\documentclass[journal]{IEEEtran}
\usepackage{graphicx}
\usepackage{amsmath}
\usepackage{subfig}
\usepackage{soul}
\usepackage{cite}
\usepackage[pdftex,dvipsnames]{xcolor}
\usepackage{layouts}
\usepackage[colorinlistoftodos,prependcaption,textsize=tiny]{todonotes}
\usepackage{xargs} 

\usepackage{color, colortbl}
\newcommandx{\improvement}[2][1=]{\todo[linecolor=Plum,backgroundcolor=Plum!25,bordercolor=Plum,#1]{#2}}
\newcommandx{\info}[2][1=]{\todo[linecolor=OliveGreen,backgroundcolor=OliveGreen!25,bordercolor=OliveGreen,#1]{#2}}

\definecolor{lg2}{rgb}{0.92,0.92,0.92}

\usepackage{bm}
\usepackage{multirow}
\begin{document}

\title{{Physics-Based Modeling and Validation of 2D Schottky Barrier Field-Effect Transistors}}

\author{\IEEEauthorblockN{Ashwin~Tunga$^{a}$,
        Zijing~Zhao,
        Ankit~Shukla,
        Wenjuan~Zhu,
        and~Shaloo~Rakheja}\\
        \IEEEauthorblockA{\textit{Holonyak Micro and Nanotechnology Laboratory}, University of Illinois at Urbana-Champaign, Urbana, IL USA}\\
        $^a$tunga2@illinois.edu}%

\maketitle

\begin{abstract}
In this work, we describe the charge transport in {two-dimensional (2D)}
Schottky barrier field-effect transistors (SB-FETs) based on the carrier injection at the Schottky contacts. {
We first develop a numerical model for thermionic and field-emission processes of carrier injection that occur at a Schottky contact. The numerical model is then simplified to yield an analytic equation for current versus voltage ($I$-$V$) in the SB-FET. The lateral electric field at the junction, controlling the carrier injection, is obtained by accurately modeling the electrostatics and the tunneling barrier width.} 
Unlike previous SB-FET models that are valid for near-equilibrium conditions, this model is applicable for a broad bias range as it incorporates the pertinent physics of thermionic, thermionic field-emission, and field-emission processes from a 3D metal into a 2D semiconductor.
The $I$-$V$ model is validated against the measurement data of 2-, 3-, and 4-layer ambipolar MoTe$_2$ SB-FETs fabricated in our lab, as well as the published data of unipolar 2D SB-FETs using MoS$_2$. Finally, the model's physics is tested rigorously by comparing model-generated data against TCAD simulation data. 
\end{abstract}


\IEEEpeerreviewmaketitle

\vspace{-5pt}
\section{Introduction}
\vspace{-5pt}
\IEEEPARstart{O}{ver} the past few decades, the semiconductor industry has focused on dimensional scaling of silicon transistors based on Moore's law in order to improve their speed, performance, and efficiency~\cite{cavin2012science, 7878935}.
However, due to the short-channel effects, such as the leakage current and static power dissipation~\cite{roy2003leakage, yadav2022extensive}, in ultra-scaled transistors~\cite{pearce1985short, mendiratta2020review, khanna2016short}, dimensional scaling has been slowing down in recent years.
Several solutions to this problem have been explored, forming what is known as the ``more-than-Moore'' strategy~\cite{waldrop2016more, lemme2014graphene}. To that end, novel materials that can mitigate short-channel effects have been investigated~\cite{wang2018low, ahmad2021evolution}. Among various novel materials, two-dimensional (2D) semiconductors have emerged as an excellent channel material for a field-effect transistor (FET)~\cite{liu2018two}. 
In a 2D semiconductor FET, mobile electrons, confined in an atomically thin channel, are strongly electrostatically coupled to the gate~\cite{chhowallaTwodimensionalSemiconductorsTransistors2016}. 
The primary advantage of 2D semiconductor FETs over ultra-thin body (UTB) transistors~\cite{low2016ultimate}
is that UTB semiconductors are a result of the termination of a 3D crystal, which leads to surface roughness and considerable carrier scatterings.
In contrast, 2D semiconductors are inherently atomically thin and do not have dangling bonds, 
and could offer higher performance at ultra-scaled process nodes. To harness the full potential of 2D materials for nanoscale CMOS, challenges related to device scaling, low resistance contacts, gate-stack design, wafer-scale integration, and process variability must be addressed. A review of opportunities and challenges of 2D semiconductors can be consulted in recent publications~\cite{knobloch2022challenges, wang2021road}.

Transition metal dichalcogenides (TMDs) are a class of 2D materials that can be incorporated into FET device structures. 
TMDs have a sizeable bandgap, transitioning from indirect bandgap in bulk to direct bandgap in their monolayer limit \cite{kumarElectronicStructureTransition2012a}. Their sizable bandgap lends to their advantage over graphene for logic devices 
~\cite{xia2010graphene}. TMD-based FETs are also expected to be superior to black-phosphorus-based FETs in which the on-off ratio degrades rapidly at high drain-bias~\cite{haratipour2016fundamental}, thus limiting the prospects of black phosphorus for low-power logic operations.  
Among the various TMD materials, MoTe$_2$ is an excellent candidate for implementing logic FETs. 
MoTe$_2$ FETs with an on-off current ratio of $10^6$ and ambipolar conduction have been experimentally demonstrated \cite{linAmbipolarMoTe2TransistorsTheir2014, larentisReconfigurableComplementaryMonolayer2017a, zhao2021nonvolatile}. 
Ambipolar transistors could reduce the complexity, while also enhancing the security~\cite{wu2021two}, of CMOS circuits since the channel can be tuned to conduct both electrons and holes by applying an appropriate electric field. 

To enable circuit design and allow technology-to-circuit co-optimization, a compact device model that faithfully reproduces the device terminal behavior over a broad operating range is needed. In the case of MoTe$_2$ SB-FETs, a physically accurate and scalable compact model must accurately interpret the role of source and drain contacts. As a result of the lack of 
effective substitutional doping techniques, metal contacts are directly deposited over the MoTe$_2$ channel~\cite{wangSchottkyBarrierHeights2021a}. Thus, unlike metal-oxide-semiconductor (MOS) FETs with ohmic source and drain contacts, MoTe$_2$ FETs invariably have Schottky contacts, which limit the injection of mobile carriers into the channel and thus the net current flow in the transistor. 

The compact model presented here is based on the generalized theory of carrier 
emission at the metal source/drain contacts. Our model includes both thermionic emission and thermal field emission (tunneling) of carriers at the Schottky contacts. Due to their closed-form nature, the equations we arrive at are easily adaptable to a compact model, where majority of the model parameters have a well-defined physical interpretation. 
We show that the model captures the essential physics of 2D SB-FETs by comparing model output against numerical simulations conducted in a commercial TCAD tool. We demonstrate the model's applicability to fabricated MoTe$_2$ ambipolar FETs with varying channel thickness (2-layers to 4-layers) and over broad bias conditions ($V_\mathrm{GS}$ $\in [-8, 8]$~V and $V_\mathrm{DS}$ $\in [0.5, 4.5]$~V), spanning both hole and electron conduction regimes. The model is also successfully applied to fabricated unipolar 2D FETs based on MoS$_2$.  

\vspace{-10pt}
\section{Prior Work}
\vspace{-5pt}
Penumatcha et al. \cite{penumatchaAnalysingBlackPhosphorus2015} developed an analytical method to describe the off-state transfer characteristics of low-dimensional FETs. The authors model the transmission of carriers through a Schottky barrier using the Landauer formalism. However, the model involves numerical computation and is thus not suitable for compact modeling. Besides, the model was demonstrated only for below-threshold gate bias $(V_\mathrm{GS})$ and low drain voltages ($V_\mathrm{DS}$), which limits the model use for practical operating conditions. Prior works also include the 
2D Pao-Sah model with drift-diffusion formalism extended to ambipolar transport \cite{guoStudyAmbipolarBehavior2016, marinNewHolisticModel2018, yarmoghaddamPhysicsBasedCompactModel2020}. These models do not account for Schottky contact-limited charge injection and instead focus on the channel-controlled charge transport, which is not the main transport physics here.
The model for SB-FETs presented in~\cite{roemerPhysicsBasedDCCompact2022}, validated only against TCAD data, is strictly derived for carrier injection into a 3D semiconductor and is thus not applicable to the 2D SB-FETs presented here. Moreover,~\cite{roemerPhysicsBasedDCCompact2022} also neglects the thermionic field emission current, which as we discuss in Sec.~\ref{sec:compact_model} is crucial for intermediate gate voltage regimes. Neglecting thermionic field emission is also expected to yield an unphysical temperature dependence of $I$-$V$ curves of an SB-FET. In~\cite{zhangSurfacePotentialCurrent2015}, a tunneling equation, empirically derived from the 3D thermionic
emission equation, along with the drift-diffusion formalism is used to obtain the drain
current in a Si nanowire FET. However, because of its implicit nature, the model is not considered compact from a circuit simulation standpoint. A compact model for a double-gated reconfigurable FET is presented in~\cite{niPhysicBasedExplicitCompact2021} based on 3D band-to-band tunneling current in an SB-FET, which is not the relevant physics underlying the ambipolar 2D SB-FETs discussed in our work. In~\cite{leeDemonstrationReconfigurableFET2022}, authors focus on the experimental demonstration of reconfigurable logic gates based on the SOI technology. On the modeling front, the authors use an empirical formulation, based on tan-hyperbolic functions to fit the experimental data. Other related works either focus on dual-gate nanowire geometry~\cite{zhu2009compact}, silicon-on-insulator structure~\cite{schwarz2013compact}, consider only 3D channels~\cite{balaguer2011analytical}, or implement a numerical $I$-$V$ model~\cite{vega2006schottky} for SB-FETs.

The model presented in this paper is specifically developed for SB-FETs using 2D semiconductors, has a strong physical basis, is validated rigorously against numerical simulations as well as experimental data of ambipolar SB-FETs fabricated in-house and unipolar SB-FETs reported in the literature. Because of its explicit nature with few parameters, most of which have a physical origin, our model is suitable for circuit simulations.

\vspace{-12pt}
\section{Model Description} \label{sec:overview}
\begin{figure}[t!]
    \centering
    \vspace{-10pt}
    \includegraphics[width=1.8in]{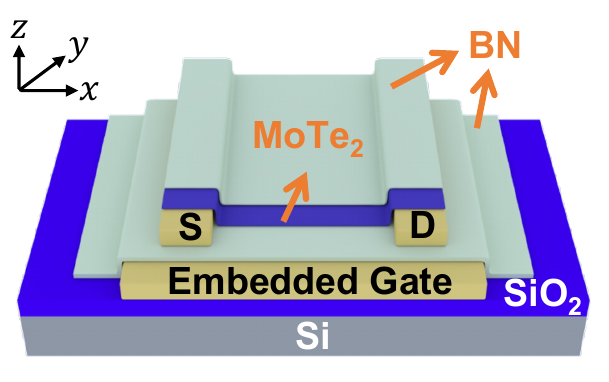}
    \vspace{-5pt}
    \caption{A schematic of the fabricated and modeled device with BN dielectric, MoTe$_2$ channel, and embedded gate. }
    \label{fig:dev_cross_sec}
\end{figure}

\vspace{-5pt}
Figure~\ref{fig:dev_cross_sec} shows the cross-section of an MoTe$_2$ SB-FET with hexagonal BN gate dielectric and metal source and drain contacts, which create a Schottky barrier at the metal/2D channel interface. 
Here, a van der Waals gap is formed between the metal and the semiconductor, resulting in a tunneling barrier, 
which increases the net contact resistance~\cite{angPhysicsElectronEmission2021}. Due to the atomic thickness of the 2D channel, the charge injection mechanism differs significantly from injection from a metal into bulk materials. Although Richardson-Dushman~\cite{richardson1925thermionic} and Fowler-Nordheim~\cite{fowler1928electron, forbes2008physics} theories of electron emission formulated for bulk materials~\cite{millikan1928relations, jensen2006general} can fit experimental data for 2D devices, these models do not represent the essential physics of 2D SB-FETs. 

\begin{figure}[t!]
\vspace{-5pt}
    \centering
    \includegraphics[width=3.5in]{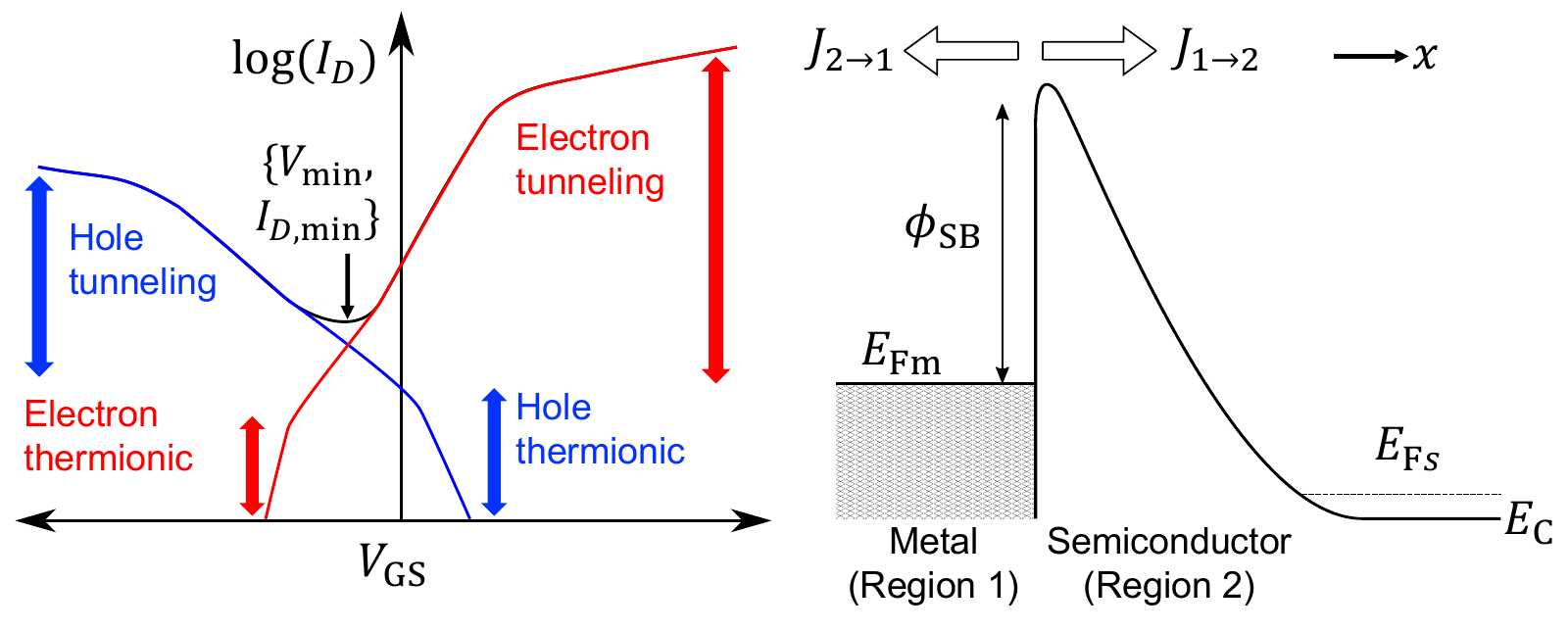}
    \caption{(left) A qualitative representation of the ambipolar SB-FET's transfer curve for an arbitrary drain bias. 
    ($\mathrm{V_{min}}$, $\mathrm{I_{D,min}}$) represents the point at which the total drain current transitions from hole-dominated to electron-dominated. (right) Energy band diagram showing carrier transmission. Here, $J$ is the current density, $\phi_\mathrm{SB}$ is the height of the Schottky barrier, $E_\mathrm{Fm}$ is the metal Fermi level, $E_c$ and $E_\mathrm{Fs}$ are the conduction band and Fermi level, respectively, in the semiconductor.}
    \label{fig:MS_barrier}
    \vspace{-15pt}
\end{figure}

In the thermionic emission (TE) process, thermally excited carriers with energy greater than the potential barrier at the contacts can traverse over the barrier into the semiconducting channel, resulting in a current flow. The activation energy for TE, \emph{i.e.}, the barrier between the metal and the bottom of the conduction band for electrons and the top of the valence band for holes, decreases linearly with gate bias until it equals the characteristic Schottky barrier height. Due to the linear variation of the activation energy, the channel current varies exponentially with gate bias, as shown in Sec.~\ref{sec:numerical_model}. 
With increasing gate bias, the potential barrier thins, which increases the probability of carriers to tunnel through the barrier. Thus, a field-dependent tunneling current is observed in the device.
The electric field-enhanced tunneling phenomenon is also referred to as field emission (FE). The sum of TE and FE currents gives the net drain current measured in an SB-FET, illustrated qualitatively in Fig.~\ref{fig:MS_barrier}(left).
Unlike in a unipolar device, in an ambipolar SB-FET, 
the TE current is marginal compared to the FE current, and the total drain current is predominantly the sum of the electron and hole FE currents.

\vspace{-15pt}
\subsection{Numerical Model} \label{sec:numerical_model}
\vspace{-5pt}
The current density, $J_\mathrm{net}$, due to carrier transmission across an energy barrier from a metal into the channel is given as $ J_{\mathrm{net}} = J_{1 \rightarrow 2} - J_{2 \rightarrow 1}$,
where $J_{1 \rightarrow 2}$ ($J_{2 \rightarrow 1}$) is the current density due to carriers incident from region $1$ (region 2) into region 2 (region 1), shown in Fig.~\ref{fig:MS_barrier}(right).
Consider $J_{1 \rightarrow 2}$:
\begin{equation}
\vspace{-10pt}
    J_{1 \rightarrow 2} = \frac{1}{\mathcal{A}}\sum_{\bm{k}} q T(k_x) v_x f_1(\bm{k}) (1-f_2(\bm{k})),
    \label{eq:Jnet1}
\end{equation}
where $q$ is the charge of the carrier, $\mathcal{A}$ is the area of the 2D crystal, $\bm{k}$ is the wavevector in reciprocal space, $k_x$ is the x-component of $\bm{k}$, $v_x$ is the velocity of carrier incident at the barrier, $f_i$ is the Fermi-Dirac distribution in region $i$ ($i=1,2$ for metal, semiconductor), and $T(k_x)$ is the transmission probability. If the carriers considered are electrons, converting the sum over $\bm{k}$-space into an integral in energy space gives
\begin{multline}
\vspace{-10pt}
        J_{1 \rightarrow 2} = -\frac{4q}{h^2} \sqrt{\frac{m_e^*}{2}}  \int_{-\infty}^\infty T(E_x) \times \\ \left( \int_{0}^\infty \frac{f_1(E) (1 - f_2(E))}{\sqrt{E_y}} \, dE_y \right) \, dE_x,
        \label{eq:J12pre}
\end{multline}
where $m_e^*$ is the effective mass of the electron, $h$ is Planck's constant, $E_x$ is the energy due to momentum perpendicular to the barrier or the longitudinal momentum (\emph{i.e.}, $E_x = p_x^2/2m_e^*$, where $p_x$ is the momentum perpendicular to the barrier interface), $E_y$ is the energy due to lateral momentum. 
The metal Fermi-level ($E_\mathrm{Fm}$) is considered as the reference energy level. The model assumes conservation of lateral carrier momentum with effective mass approximation.
 
The same procedure as above can be followed to obtain an expression for $J_{2 \rightarrow 1}$ to get $J_{\mathrm{net},e}$ as
 \begin{multline} \label{eq:jnet_e}
     J_{\mathrm{net},e} = -\frac{4q}{h^2} \sqrt{\frac{m_e^*}{2}}  \int_{-\infty}^\infty T(E_x) \times \\ \left( \int_{0}^\infty \frac{f_1(E) - f_2(E)}{\sqrt{E_y}} \, dE_y \right) \, dE_x.
 \end{multline}
The net hole current density is obtained similarly as the electron current with $m_e^*$ replaced with the effective mass of holes, $m_h^*$.
The total drain current density due to both carriers is simply the sum of their respective net currents. Taking into account the direction of the drain current (along the -$x$ axis, from drain to source contact), the total current density is $ -J_{\mathrm{D}} =J_{\mathrm{net},e} + J_{\mathrm{net},h}$.
While the drain current is modeled by considering the spatially localized carrier injection at the contacts, effects of gate-source voltage ($V_\mathrm{GS}$) and drain-source voltage ($V_\mathrm{DS}$) 
are incorporated via the Fermi functions, $f_1$ and $f_2$, and the transmission probability, $T(E_x)$.

For the classical TE process, the transmission probability $T(E_x) = 1$. Evaluating (\ref{eq:jnet_e}) for electrons for $T(E_x) = 1$ and non-degenerate statistics and integrating over $E_x \in [E_{A}, \infty)$ ($E_A$ is the TE activation energy) gives 
(see Appendix \ref{app:thermionic})
\begin{multline} \label{eq:thermionic_e}
\vspace{-10pt}
    J_{\mathrm{TE,}e} = -q \frac{\sqrt{8\pi k_B^3 m_e^*}}{h^2} T^{3/2} \exp \left(-\frac{E_{A}}{k_B T} \right) \\
    \times \left[ 1 - \exp \left( -\frac{qV_c}{k_B T} \right) \right].
\end{multline}
The activation energy $E_{A}$ reduces linearly with the gate bias until the flatband voltage when $E_{A} = \phi_\mathrm{SB}$, $qV_c = E_\mathrm{Fm} - E_\mathrm{Fs} = - E_\mathrm{Fs}$ is the voltage across the contact ($E_\mathrm{Fm}$ is used as the reference energy level), 
$A^*_{2D} \equiv (q\sqrt{8\pi k_B^3 m_e^*})/(h^2)$ is the effective Richardson constant for a 2D semiconductor. It is important to note the $T^{3/2}$ dependence in the pre-factor of the above equation compared to the $T^2$ dependence in the 3D TE model~\cite{szePhysicsSemiconductorDevices2021}. A similar treatment leads to the hole TE current. 

The FE transmission probability using the Wentzel–Kramers–Brillouin (WKB) approximation for a triangular barrier~\cite{schwarz2013compact} is
  \begin{equation}\label{eq:wkb_tunn_prob}
     T_{e,h}(E_x) = \exp{\left( -\frac{8 \pi \sqrt{2{m^*} (\phi_{\mathrm{SB},(e,h)} - E_x)^3}}{3hqF_x} \right)},
 \end{equation}
where $F_x$ is the magnitude of the electric field at the triangular barrier. 
For electron tunneling that dominates for $V_\mathrm{GS}>V_\mathrm{min}$, (\ref{eq:jnet_e}) is integrated over $E_x \in [-\infty,\phi_\mathrm{SB}]$.
In Sec.~\ref{sec:compact_model}, we show an explicit analytic tunneling equation that lends itself well to compact modelling. 

\begin{figure}[t!]
    \centering
    \vspace{-10pt}
    \includegraphics[width=0.7\linewidth]{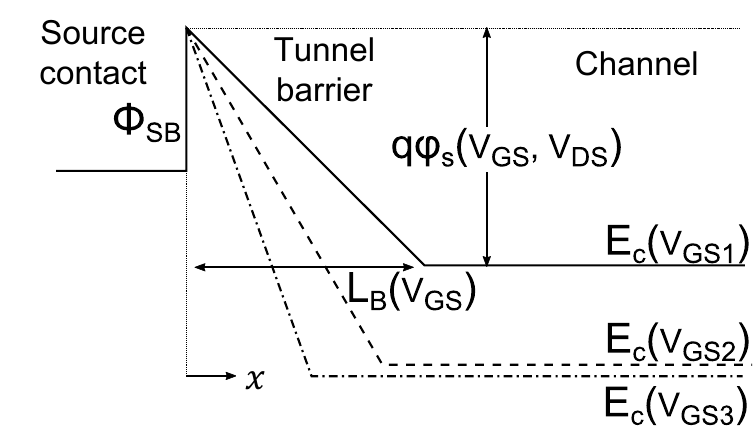}
    \caption{A simplified conduction band profile at the source for fixed $V_\mathrm{DS}$. $V_\mathrm{GS3} > V_\mathrm{GS2} > V_\mathrm{GS1}$. $\phi_\mathrm{SB}$ is the Schottky barrier height, $\varphi_s$ is the surface potential,
    $L_B$ is the tunneling barrier width. As $V_\mathrm{GS}$ increases, $\varphi_s$ increases and $L_B$ reduces, increasing $F_x$. Beyond strong inversion, $\varphi_s$ saturates while $L_B$ continues decreasing, which reduces the rate of increase of $F_x$.}
    \label{fig:Fx_vgs_rep}
    \vspace{-5pt}
\end{figure}

A simplified conduction band profile at the source contact is shown in Fig. \ref{fig:Fx_vgs_rep}. The electric field is 
given as
\begin{equation} \label{eq:F_x_emp}
    F_x = \frac{\varphi_{(s,d)}(V_\mathrm{GS}, V_\mathrm{DS})}{L_B(V_\mathrm{GS}, V_\mathrm{DS})},
\end{equation}
where $L_B$ is the tunneling barrier width, and $\varphi_{(s,d)}$ is the potential drop at the respective source/drain contact. For a constant $V_\mathrm{DS}$, as $V_\mathrm{GS}$ increases, $\varphi_s$ increases and $L_B$ decreases, resulting in a strong increase in $F_x$ with $V_\mathrm{GS}$. At yet higher $V_\mathrm{GS}$, $\varphi_s$ remains roughly constant but $L_B$ continues to decrease, which reduces the rate of increase of $F_x$ with $V_\mathrm{GS}$.
In our model, the effect of the channel transport is enclosed in the electric field, which ensures the self-consistency between our methodology and the emission-diffusion theory of MOSFETs presented in~\cite{lundstrom2015emission}.

\vspace{-15pt}
\subsection{Compact Model} \label{sec:compact_model}
\newcommand{\V}[1]{V_{#1}}
The SB-FET compact model includes both TE and FE processes and is thus applicable over a broad bias range. 
The total drain current is $ I_{D} = W\left[(J_\mathrm{tun,e} + J_\mathrm{TE,e}) + (J_\mathrm{tun,h} + J_\mathrm{TE,h})\right]$, where $W$ is the device width, and $J_\mathrm{tun}$ ($J_\mathrm{TE}$) is the tunneling (thermionic) current. The holes (electrons) are injected into the channel from the drain (source) contact.
The TE process for a 2D semiconductor is described by (\ref{eq:thermionic_e}), while 
the 2D tunneling process can be described by the following set of equations (see Appendix~\ref{app:field_emission}):
\begin{equation} \label{eq:compact_jtun}
    J_\mathrm{tun,(e,h)} = P_{f,(e,h)} \left( J_\mathrm{TFE} + J_\mathrm{FE} \right),
\end{equation}
\begin{equation}
    J_\mathrm{TFE} = C_0 C_1 \frac{\sqrt{k_BT\pi}}{\beta} \left( \exp(\beta\phi_\mathrm{SB,(e,h)}) - 1 \right),
\end{equation}
\begin{equation}
    J_\mathrm{FE} = C_0 C_1 \sqrt{\pi E_{00}^3} \left(1 - \exp\left(- \frac{qV_\mathrm{DS}}{E_\mathrm{00}} \right) \right),
\end{equation}
\begin{equation} \label{eq:compact_jtun_consts}
\begin{gathered}
    C_0 = \frac{4q}{h^2} \sqrt{\frac{m^*_{(e,h)}}{2}}, \\
    C_1 = \exp\left( - \sqrt{\phi_\mathrm{SB,(e,h)} - E_0} \left( \phi_{SB,(e,h)} + \frac{E_0}{2}  \right)  \right), \\
    E_0 = \phi_{SB,(e,h)} - \left(\frac{\ln(K_0)}{\alpha}\right)^{\frac{2}{3}}, \\
    \beta = \frac{k_BT - E_{00}}{(k_BT)(E_{00})}, \,
    E_{00} = \frac{2}{3\alpha\sqrt{\phi_{SB,(e,h)} - E_0}}, \\
    \alpha = \frac{8\pi\sqrt{2m^*_{(e,h)}}}{3hqF_{x}}.
\end{gathered}
\end{equation}
$J_\mathrm{TFE}$ and $J_\mathrm{FE}$ are the thermionic field emission (TFE) and field emission (FE) components, respectively, of the tunneling current, and $K_0$ and $P_f$ are constant fitting parameters. The terminal voltages modulate $F_x$ at the Schottky contact and thus control the current through the device.

To model $F_x$, we need to obtain the channel potential and the tunneling barrier width.
The channel potential is obtained from the balance equation given as
\begin{equation}
    \begin{gathered}
        V_\mathrm{G(S,Deff)} - V_\mathrm{FB,(e,h)} = \varphi_{(s,d)} - \frac{Q_\mathrm{ch,(e,h)}}{C_\mathrm{ins}},
    \end{gathered}
\end{equation}
where $V_\mathrm{FB}$ is the flat-band voltage, $Q_\mathrm{ch}$ is the mobile charge in the channel, and $C_\mathrm{ins}$ is the insulator capacitance. 
$Q_\mathrm{ch}$ is empirically modeled as~\cite{khakifiroozSimpleSemiempiricalShortChannel2009, rakhejaAmbipolarVirtualSourceBasedChargeCurrent2014} 
\begin{equation}
    \begin{gathered}
        Q_\mathrm{ch,e} = -C_\mathrm{inv,e} n_{e} \frac{k_B T}{q} \log \left( 1 + \exp \left( q \frac{V_\mathrm{GS} - V_{T,e}}{n_e k_B T} \right) \right), \\
      Q_\mathrm{ch,h} = C_\mathrm{inv,h} n_h \frac{k_B T}{q} \log \left( 1 + \exp \left( -q \frac{V_\mathrm{GDeff} + V_{T,h}}{n_h k_B T} \right) \right),
    \end{gathered}
\end{equation}
where $C_{\mathrm{inv}(e,h)}$ is the inversion capacitance, $V_{T(e,h)}$ is the threshold voltage, and $n_{(e,h)}$ is related to the sub-threshold swing.
The $V_\mathrm{DS}$ dependence of $V_\mathrm{FB}$ and $V_T$ is given as 
\begin{equation}
    \begin{gathered}
        V_\mathrm{FB} = V_\mathrm{FB0} - \delta_\mathrm{FB}V_\mathrm{DSeff},\\
        V_{T} = V_{T0} - \delta_T \sqrt{V_\mathrm{DSeff}},
    \end{gathered}
\end{equation}
where $\delta_\mathrm{FB}$ and $\delta_T$ are empirical parameters that are determined from calibrating the model with experimental data, as described in the companion paper. $V_\mathrm{DSeff}$ is the effective $V_\mathrm{DS}$ that drops across the channel. The effective $V_\mathrm{DS}$ varies linearly with $V_\mathrm{DS}$ at low drain bias and eventually saturates at $V_\mathrm{DSAT}$. We define $V_\mathrm{DSeff}$ and $V_\mathrm{GDeff}$ using a saturation function as,
\begin{equation}
    \begin{gathered}
        V_\mathrm{DSeff} = V_\mathrm{DS} \frac{V_\mathrm{DS}/V_\mathrm{DSAT}}{\Big( 1 + (V_\mathrm{DS}/V_\mathrm{DSAT})^\nu \Big)^\frac{1}{\nu}}, \\
        V_\mathrm{GDeff} = V_\mathrm{GS} - V_\mathrm{DSeff}.
    \end{gathered}
\end{equation}
Here, $V_\mathrm{DSAT}$ is the saturation voltage and $\nu$ is the transition region fitting parameter.

\begin{figure}[t!]
\vspace{-12pt}
    \centering
    \includegraphics[width=2.60in]{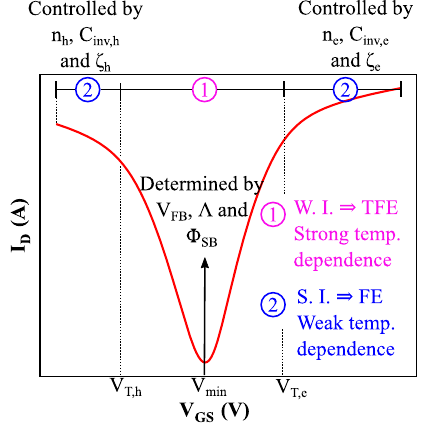}
   \vspace{-5pt}
    \caption{A representative model-generated $I_D$-$V_\mathrm{GS}$ curve. W.I. (S.I.) stands for weak (strong) inversion.}
    \label{fig:model_rep_idvg}
    \vspace{-10pt}
\end{figure}

The tunneling barrier width,
$L_B$, depends on the 
characteristic length, $\lambda$, and the depletion width, $W_D$. The tunneling process can happen either over $W_D$ or a few characteristic lengths ($\Lambda = n_0 \lambda$, $n_0 \geq 1$). 
At low $V_\mathrm{GS}$, $W_D$ is greater than $\Lambda$, and $L_B$ is determined by $\Lambda$. At intermediate $V_{GS}$, 
the depletion region thins and $W_D<\Lambda$,
and the tunneling path is influenced by the depletion width. 
Thus, $L_B$ is modeled as
\begin{equation}
\begin{gathered}
L_B = \frac{\Lambda W_D}{\Lambda + W_D}, \text{ }
    \Lambda = n_0 \lambda = n_0 \sqrt{\frac{t_\mathrm{ch}t_\mathrm{ins} \epsilon_\mathrm{ch}}{\epsilon_\mathrm{ins}}}, \\
    W_{D,(e,h)} = \sqrt{\frac{\epsilon_\mathrm{ch} \varphi_{(s,d)}}{\zeta_{(e,h)} Q_\mathrm{ch,(e,h)}/t_\mathrm{ch}}},
\end{gathered}
\end{equation}
$t_\mathrm{ch}$ ($t_\mathrm{ins}$) is the channel (insulator) thickness, 
$\epsilon_\mathrm{ch}$ ($\epsilon_\mathrm{ins}$) is the channel (insulator) dielectric constant, and $\zeta$ is a fitting parameter that describes the charge in the depletion region as a fraction of the channel charge. 
Figure~\ref{fig:model_rep_idvg} shows the effect of key model parameters on a typical 
$I_D$-$V_\mathrm{GS}$ curve.

\begin{figure*}[t!]
\vspace{-12pt}
    \centering
    \includegraphics[width=7in]{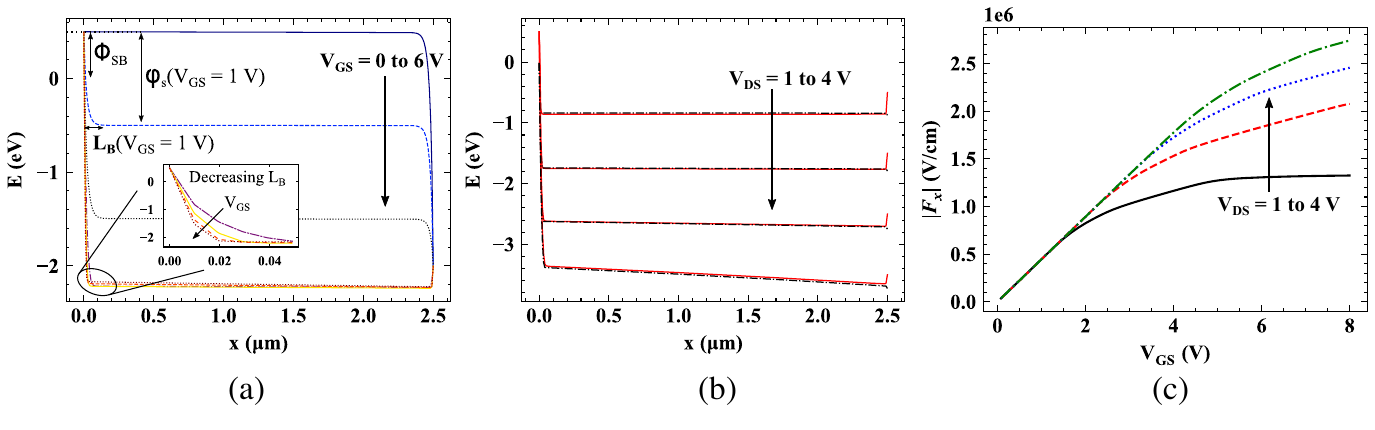}
    \vspace{-12pt}
    \caption{(a) Conduction band profile along the channel length at $V_\mathrm{DS} = 2.5$~V for various $V_\mathrm{GS}$ values. $\phi_\mathrm{SB}$ is the Schottky barrier height, while $\varphi_s$ is the potential drop across the source contact. As $V_\mathrm{GS}$ is increased, $\varphi_s$ saturates. The inset figure shows the band profile magnified near the source contact region. 
    (b) Conduction band profile along the channel at $V_\mathrm{GS} = 5$~V for various applied drain bias. The solid lines represent the conduction band and the dash-dotted line represent the electron quasi-Fermi level. Here, electron tunneling dominates. (c) $F_x$ vs $V_\mathrm{GS}$ near the source for a four-layer device.}
    \vspace{-12pt}
    \label{fig:tcad_validation}
\end{figure*}

\vspace{-10pt}
\section{Model validation} \label{sec:model_val}
\subsection{Comparison against TCAD results}
The device physics of MoTe$_2$ SB-FETs is analyzed using the TCAD tool, Sentaurus, from Synopsys \cite{SentaurusDeviceUser2020}.  
A four-layer, 2.5 ~$\mathrm{\mu m}$ long MoTe$_2$ SB-FET with 30~nm thick BN gate dielectric was simulated.
The band-gap of MoTe$_2$ was fixed at 1.0~eV, while the hole and electron effective masses were kept equal at $0.55m_0$ ($m_0$ is the free electron mass.). 
Further, the source/drain contacts were modeled as Schottky contacts with a Schottky barrier height of $0.50$~eV. Finally, the gate contact was treated as a Dirichlet boundary condition, and the rest of the boundaries were treated as a Neumann boundary.

The drift-diffusion formalism was used to model the 
charge transport in the channel, 
with carrier mobility fixed at $50$~cm$^2$/Vs for both electrons and holes. Injection at the source and drain contacts was modeled using 
thermionic emission and the non-local tunneling equations, as implemented in Sentaurus.
TCAD simulation results, shown in Fig.~\ref{fig:tcad_validation}(a), confirm that the 
the majority of $V_\mathrm{DS}$ drops at the contacts. 
Moreover, from Fig.~\ref{fig:tcad_validation}(b), we can infer that the 
charge transport is severely limited by carrier injection at the contacts and that the region near the contacts is depleted of charge carriers.

The dependence of the electric field, $F_x$, at the source contact with $V_\mathrm{GS}$ is shown in Fig. \ref{fig:tcad_validation}(c). $F_x$, which is given as the ratio of the potential drop, $\varphi_s$ at the source and the depletion width, $L_B$, increases linearly with $V_\mathrm{GS}$ in weak inversion.
This is because in weak inversion 
$\varphi_s$ varies linearly with $V_\mathrm{GS}$, while $L_B$ remains constant. 
At high $V_\mathrm{GS}$, although $L_B$ continues to shrink as shown in Fig.~\ref{fig:tcad_validation}(b), $\varphi_s$ saturates, which slows the rate of increase of $F_x$ in strong inversion.

\vspace{-14pt}
\subsection{Comparison against measurement data}
\begin{figure}[t!]
\vspace{-5pt}
    \centering
    \includegraphics[width=1.9in]{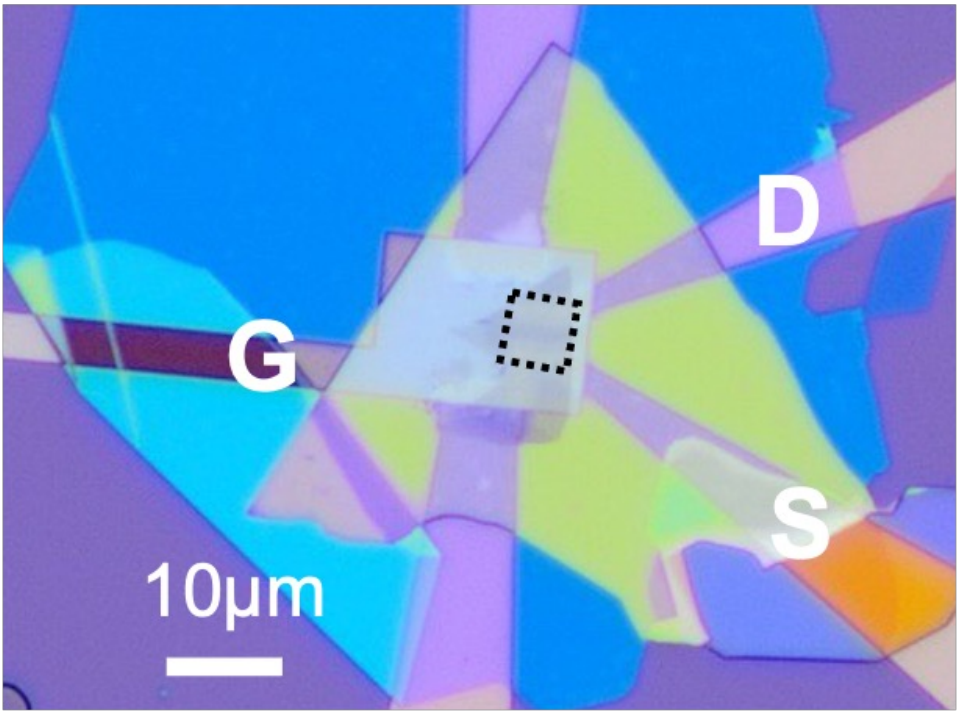}
    \vspace{-2pt}
    \caption{Optical image of a trilayer MoTe$_2$ SB-FET. The MoTe$_2$ channel is encapsulated by bottom (yellow) and top (blue) BN layers. The MoTe$_2$ channel is highlighted in the dashed box.}
    \label{fig:optical}
    \vspace{-15pt}
\end{figure}

\vspace{-5pt}
We validate our compact model against experimental measurement data of bilayer, trilayer, and four-layer MoTe$_2$ SB-FETs fabricated in-house. Figure~\ref{fig:optical} shows the optical image of a fabricated trilayer MoTe$_2$ device. The devices were fabricated by following a bottom-up approach, where the embedded gates are formed first with metal evaporation after optical lithography patterning. The bottom BN was exfoliated and transferred on to the gates using the dry transfer method, before the source and drain contacts were patterned. The MoTe$_2$ flakes were exfoliated on to a 90~nm SiO$_2$/Si wafer. The thicknesses were identified from the optical image contrast. The MoTe$_2$ flakes are dry transferred on top of the source and drain contacts, with a top BN flake as the adhesive layer. The top BN layer also encapsulates the MoTe$_2$ channel. 

\begin{figure*}[t!]
    \centering
    \includegraphics[width=7in]{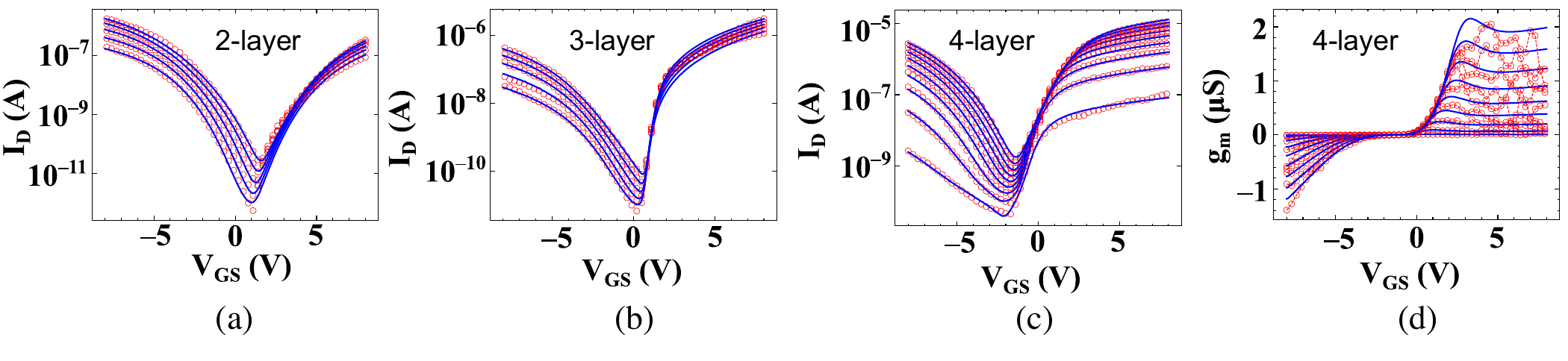}
    \vspace{-12pt}
    \caption{(a)-(c) Transfer characteristics of various MoTe$_2$ SB-FETs. (d) Transconductance of the 4-layer Mote$_2$ SB-FET. In each case, the drain voltage varies from 2.5 V (bottom-most trace) to 4.5 V (top-most trace) in steps of 0.5 V.}
    \label{fig:expt_calibrate}
\end{figure*}

Our model contains a total of 24 parameters (11 each for electrons and holes and 2 common to both). Nine of the parameters are empirical in nature, while the remainder have a physical origin and can be deduced from straightforward experimental calibration. See Appendix~\ref{app:par_extract_method} for parameter extraction methodology.
Figure~\ref{fig:expt_calibrate}
shows an excellent match between the transfer curves obtained from our model and measurement data of the bilayer, trilayer, and four-layer MoTe$_2$ SB-FETs. Additionally, our model can capture the transconductance of the device measured experimentally.
Table~\ref{tab:model_param} shows the extracted model parameters.
The asymmetric electron and hole conduction in the fabricated devices is due to the unequal Schottky barrier heights, 
$\phi_{\mathrm{SB},e} \neq \phi_{\mathrm{SB},h}$.
The extracted values of $\phi_{\mathrm{SB},e}$ and $\phi_{\mathrm{SB},h}$
show that $E_g (= \phi_{\mathrm{SB},e} + \phi_{\mathrm{SB},h})$ increases with the increase in the channel thickness
~\cite{ruppertOpticalPropertiesBand2014, lezamaIndirecttoDirectBandGap2015}.

\begin{figure}[h!]
    \centering
    \vspace{-10pt}
    \includegraphics[width=0.75\linewidth]{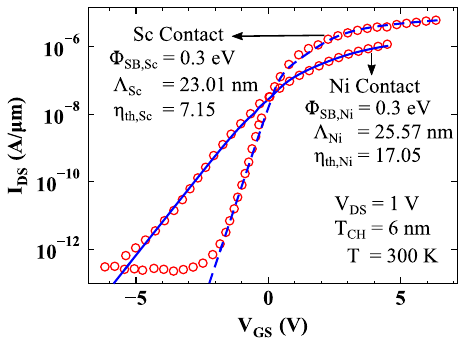}
    \vspace{-5pt}
    \caption{Transfer curves of 6-nm-thick MoS$_2$ transistors with Ni and Sc contacts. Experiments~\cite{dasHighPerformanceMultilayer2013}: symbols; Model: lines.}
    \label{fig:unipolar_idvg}
    \vspace{-10pt}
\end{figure}

We also apply our compact model to n-type MoS$_2$ SB-FETs reported in~\cite{dasHighPerformanceMultilayer2013}.
In a unipolar device, TE is observable in the measured $I-V$ data in the sub-threshold regime. This is readily captured in our model as it is based on a generalized theory of carrier emission.
Figure~\ref{fig:unipolar_idvg} shows that the model faithfully captures the channel current from sub-threshold to strong inversion regimes for a 6-nm thick and 5-$\mu$m long MoS$_2$ FET. At very low gate voltages, the Sc contacted device is dominated by gate leakage, which is not included in our model.

\begin{table*}[]
\vspace{-5pt}
    \centering
    \renewcommand{\arraystretch}{1.1}
        \caption{Extracted parameters for the device model. For all cases, $m_e^* = 0.57m_0$, $m_h^* = 0.75m_0$~\cite{ilatikhamenehTunnelFieldEffectTransistors2015}. $\epsilon_\mathrm{ch}=8.6$ for bilayer and trilayer MoTe$_2$, while $\epsilon_\mathrm{ch}=11.2$ for four-layer MoTe$_2$~\cite{kumarTunableDielectricResponse2012}, $\epsilon_\mathrm{ins}=3.8$~\cite{laturiaDielectricPropertiesHexagonal2018a}. Gray cells: Purely empirical parameters.}
        \vspace{-5pt}
    \begin{tabular}{|c|c|c|c|c|c|c|c|}
        \hline
        \multicolumn{2}{|c|}{\textbf{Model Parameter}} &  \multicolumn{2}{c|}{\textbf{Bilayer}} & \multicolumn{2}{c|}{\textbf{Trilayer}} & \multicolumn{2}{c|}{\textbf{Four-layer}} \\
        \hline
         \textbf{Description} & \textbf{Symbol}  & \textbf{Electrons} & \textbf{Holes} & \textbf{Electrons} & \textbf{Holes} & \textbf{Electrons} & \textbf{Holes} \\
         \hline
         Schottky barrier height [eV] & $\phi_\mathrm{SB}$ & 0.62 & 0.59 & 0.56 & 0.54 & 0.53 & 0.48\\ \hline
         Inversion capacitance [mF/m$^2$] & $C_\mathrm{inv}$ & 1.20 & 1.13 & 1.83 & 1.79 & 1.93 & 1.93 \\ \hline
         Flatband voltage [V] & $V_\mathrm{FB0}$ & -0.40 & 0.44 & -0.28 & 0.94 & -3.17 & -2.79 \\ \hline
         Drain bias dependent flatband voltage variation & $\delta_\mathrm{FB}$ & 0.6 & -0.052 & 0.14 & -0.13 & -0.076 & -0.878\\ \hline
         Threshold voltage [V] & $V_{T0}$ & -0.30 & -6.76 & 0.58 & -7.59 & -2.52 & 1.98 \\ \hline
         Drain bias dependent threshold voltage variation [$\mathrm{V^{1/2}}$] & $\delta_{T}$ & 2.5 & 8.08 & 0.16 & 8.06 & 2.53 & 4.66 \\ \hline
         Characteristic tunneling barrier length [nm] & $\Lambda$ & 55.00 & 38.11 & 18.09 & 61.68 & 20.92 & 46.67 \\ \hline
        Saturation drain voltage [V] & $V_\mathrm{{DSAT}}$ & \multicolumn{2}{c|}{4.90} & \multicolumn{2}{c|}{4.41} & \multicolumn{2}{c|}{3.93} \\ \hline
         \rowcolor{lg2} Saturation transition parameter & $\nu$ & \multicolumn{2}{c|}{17.10} & \multicolumn{2}{c|}{9.89}  & \multicolumn{2}{c|}{3.76} \\ \hline
       \rowcolor{lg2}  Empirical channel charge modeling parameter & $n$ & 60.00 & 60.41 & 6.31 & 49.70 & 37.25 & 49.98 \\ \hline
       \rowcolor{lg2}  Empirical parameter modeling the depletion charge & $\zeta$ & 0.089 & 0.012 & 0.21 & 0.01 & 0.01 & 0.004 \\ \hline
       \rowcolor{lg2}  Empirical fitting parameter used to derive analytic tunneling equation & $K_{0}$& 100 & 100 & 100 & 100 & 100 & 100 \\ \hline
      \rowcolor{lg2}   Empirical fitting parameter to fit analytic equation with numerical equation & $P_{f}$& 0.1 & 0.3 & 0.3 & 0.1 & 0.5 & 0.5\\ \hline
    \end{tabular}
    \label{tab:model_param}
\end{table*}

\vspace{-10pt}
\section{Conclusion}
\vspace{-5pt}
A compact model for ambipolar MoTe$_2$ SB-FETs was presented. The model relies on explicit, analytic equations to model thermionic emission and field-emission tunneling. We also presented a model for the variation of the tunneling barrier width with the terminal voltages. We conducted TCAD simulations to verify the model physics. Finally, we demonstrated the model's applicability to produce $I$-$V$ data of realistic devices by comparing the model output against measurements of SB-FETs fabricated in-house as well as data available in the published literature. Because of its compact nature and few parameters, most of which have a physical significance, the model is suitable for technology-device-circuit co-design.
\vspace{-10pt}

\appendices
\renewcommand{\theequation}{A.\arabic{equation}}
\setcounter{equation}{0}
\vspace{-5pt}
\section{Derivation of thermionic emission equation} \label{app:thermionic}
\vspace{-5pt}
To derive (\ref{eq:thermionic_e}), we assume $T(E_x) = 1$ and apply Maxwell-Boltzmann statistics~\cite{li2022unified} in (\ref{eq:jnet_e}) along with $E=E_x+E_y$.
 \begin{multline}
     J_{\mathrm{net},e} = -\frac{4q}{h^2} \sqrt{\frac{m_e^*}{2}}  \int_{0}^\infty T(E_x) \times \\ \left( \int_{0}^\infty \frac{f_1(E) - f_2(E)}{\sqrt{E_y}} \, dE_y \right) \, dE_x.
 \end{multline}
 \begin{multline}
     J_{\mathrm{net},e} = -\frac{4q}{h^2} \sqrt{\frac{m_e^*}{2}} \int_{\phi_{SB}}^\infty \int_{0}^\infty \exp\left(-\frac{E_x+E_y}{k_BT}\right) \\ - \exp\left(\frac{E_{Fs} - (E_x+E_y)}{k_BT}\right) \frac{1}{\sqrt{E_y}} \, dE_y \, dE_x.
 \end{multline}
  \begin{multline}
     J_{\mathrm{net},e} = -\frac{4q}{h^2} \sqrt{\frac{m_e^*}{2}} \int_{\phi_\mathrm{SB}}^\infty \exp\left(-\frac{E_x}{k_BT}\right) \, dE_x \\ \times \int_{0}^\infty \left(\exp\left(-\frac{E_y}{k_BT}\right) - \exp\left(\frac{E_\mathrm{Fs} - E_y}{k_BT}\right) \right) \frac{1}{\sqrt{E_y}} \, dE_y.
 \end{multline}
 Solving the integrals and using $qV_c = -E_{Fs}$,
 \begin{multline}
    J_\mathrm{TE,e} = -q \frac{\sqrt{8\pi k_B^3 m_e^*}}{h^2} T^{3/2} \exp \left(-\frac{\phi_\mathrm{SB}}{k_B T} \right) \\
    \times \left[ 1 - \exp \left( -\frac{qV_c}{k_B T} \right) \right]. 
\end{multline}
\renewcommand{\theequation}{B.\arabic{equation}}
\setcounter{equation}{0}
\vspace{-15pt}
\section{Derivation of analytic tunneling equation} \label{app:field_emission}
\begin{figure}
    \centering
    \includegraphics[width=3.5in]{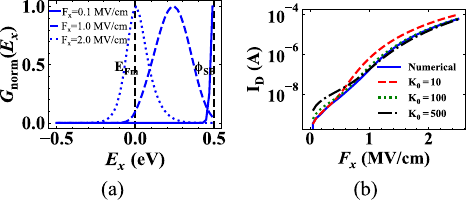}
    \vspace{-15pt}
    \caption{(a) Normalized $G(E_x)$ vs $E_x$ for various $F_x$. $E_\mathrm{Fm}$: metal Fermi level and $\phi_\mathrm{SB}$: Schottky barrier height. The peak of $G_\mathrm{norm}$ moves closer to the metal Fermi-level as $F_x$ is increased, indicating that the mechanism of carrier injection changes from TE to TFE and finally to FE as $F_x$ increases. (b) $I_D$-$F_x$ curve produced by the analytic model compared with the numerical model for different $K_0$ values.}
    \label{fig:ef_K0}
    \vspace{-15pt}
\end{figure}

FE can be analytically derived from (\ref{eq:jnet_e}) and (\ref{eq:wkb_tunn_prob}) as follows.

\begin{multline} \label{eq:app_jnet}
    J_\mathrm{tun,e} = C_0 \int_{-\infty}^{\phi_\mathrm{SB}} \exp{\left( -\alpha(\phi_{SB} - E_x)^{\frac{3}{2}} \right)} \\
                \times \left( \int_0^{\infty} \frac{f_1(E) - f_2(E)}{\sqrt{E_y}} dE_y \right) dE_x,
\end{multline}
where $J_\mathrm{tun,e}$ is the electron tunneling current, $C_0$ is the constant prefactor in (\ref{eq:jnet_e}) and $\alpha$ is the constant in the exponent in (\ref{eq:wkb_tunn_prob}).
Since $E = E_x + E_y$, the integral with respect to $E_y$ is the difference of Fermi integrals of order $-\frac{1}{2}$ given as
\begin{multline}
    \int_0^{\infty} \frac{f_1(E) - f_2(E)}{\sqrt{E_y}} dE_y = \sqrt{k_BT} \,\Gamma\left(\frac{1}{2}\right) \\
                \times \left[ \mathcal{F}_{-\frac{1}{2}}\left( \frac{E_{Fm} - E_x}{k_BT} \right) - \mathcal{F}_{-\frac{1}{2}}\left( \frac{E_{Fs} - E_x}{k_BT} \right)  \right].
\end{multline}
The Fermi integral can be approximated as
\begin{equation}
    \mathcal{F}_{-\frac{1}{2}}(x) = 
    \begin{cases}
        \exp(x),\, x \leq 0, \\
        \frac{2}{\sqrt{\pi}} x^{\frac{1}{2}},\, x>0.
    \end{cases}
\end{equation}
Equation (\ref{eq:app_jnet}) can now be converted from a double integral equation to a single integral equation in $E_x$ as follows
\begin{equation}
    J_\mathrm{tun,e} = \int_{-\infty}^{\phi_{SB}} G(E_x) dE_x,
    \label{eq:Jtunapp}
\end{equation}
where 
$G(E_x)$ is the electron tunneling current density per energy level, $E_x$, in the conduction band.
As shown in Fig. \ref{fig:ef_K0}(a), the peak of $G(E_x)$ moves closer to $E_\mathrm{Fm}$ as the electric field, $F_x$, increases. Let us define the point $E_0$ as
\begin{equation}
    E_0 = \operatorname*{arg\,max}_{E_x} G(E_x).
\end{equation} 

To obtain an analytic solution of (\ref{eq:Jtunapp}), $\ln(T(E_x))$ is linearized around $E_0$. 
Let us suppose that
\begin{equation}
    T(E_0) = \frac{1}{K_0(F_x)}.
\end{equation}
Although $K_0(F_x)$ can be approximated as a constant value, a piece-wise value of $K_0$ is a better approximation as shown in Fig.~\ref{fig:ef_K0}(b). $E_0$ is given as
\begin{equation}
    E_0 = \phi_\mathrm{SB} - \left(\frac{\ln(K_0)}{\alpha}\right)^{\frac{2}{3}},
\end{equation}
\begin{equation}
    \ln(T(E_x)) = -\alpha f(E_x) = -\alpha\left[ f(E_0) + (E_x - E_0)f'(E_0) \right].
\end{equation}
The integral in (\ref{eq:app_jnet}) now has an analytic solution.
\begin{equation}
    \begin{split}
        J_\mathrm{TFE} & = \int_0^{\phi_\mathrm{SB}} G(E_x) dE_x \\
                & = C_0 C_1 \frac{\sqrt{k_BT\pi}}{\beta} \left( \exp(\beta\phi_\mathrm{SB}) - 1 \right)\\
                &  \times \left(1 - \exp\left(- \frac{qV_{c}}{k_BT}\right) \right),
    \end{split}
\end{equation}
\begin{equation}
\begin{gathered}
    C_1 = \exp\left( - \sqrt{\phi_\mathrm{SB} - E_0} \left( \phi_\mathrm{SB} + \frac{E_0}{2}  \right)  \right), \\
    \beta = \frac{k_BT - E_\mathrm{00}}{(k_BT)(E_\mathrm{00})}, \,\,
    E_{00} = \frac{2}{3\alpha\sqrt{\phi_\mathrm{SB} - E_0}}.
\end{gathered}
\end{equation}
\begin{equation}
    \begin{split}
        J_{FE} & = \int_{-\infty}^{0} G(E_x) dE_x \\ 
               & = C_0 C_1 \sqrt{\pi E_\mathrm{00}^3} \left(1 - \exp\left(- \frac{qV_{c}}{E_{00}} \right) \right),
    \end{split}
\end{equation}
Figure \ref{fig:ef_K0}(b) shows the validation of the analytic equation with the numerical integral, using different values of $K_0$. To fit the numerical integral, another parameter is introduced for tunneling current, which gives $J_\mathrm{tun,e} = P_{f,e} (J_\mathrm{TFE} + J_\mathrm{FE})$. A similar parameter, $P_{f,h}$ is introduced for the hole branch. 

\vspace{-10pt}
\section{Parameter Extraction Methodology} \label{app:par_extract_method}
\vspace{-5pt}
The input parameters of the model that are fixed include
(i) device width ($W$) (ii) the channel thickness ($t_\mathrm{ch}$), (iii) insulator thickness ($t_\mathrm{ins}$), which along with the insulator dielectric constant ($\epsilon_\mathrm{ins}$), gives the insulator capacitance ($C_\mathrm{ins}$).

The approximate range of Schottky barrier heights ($\phi_\mathrm{SB}$) can be obtained by extracting the $x$-direction electric field at the contact ($F_x$) from the measurement data, for a given $\phi_\mathrm{SB}$, using (8)-(11) and verifying that the extracted $F_x$ is reasonable. $\phi_\mathrm{SB}$ can then be tuned to obtain a best fit. If the thermionic emission current is observed in the device, $\phi_\mathrm{SB}$ can also be extracted using the Arrhenius plots.

The minimum current points are used to determine $V_{\mathrm{FB0},(e,h)}$ and $\delta_{\mathrm{FB},(e,h)}$. The knee point in the semi-log $I_\mathrm{D}$~-~$V_\mathrm{G}$ curve determines $V_{T0}$ and $\delta_{T}$, and $n$ is related to the sharpness of the knee point. The slope of the semi-log transfer curve in the sub-threshold region are used to obtain $\Lambda$, while $\zeta$ is correlated to the on-state current of the device.

The empirical parameter $K_0$ is used to obtain an analytic tunneling equation from the numerical model. 
Lower (higher) $K_0$ approximates the low-field (high-field) region better. $K_0 = 100$ reasonably approximates tunneling at both low-field and high-field region. The empirical parameter, $P_f$, lies in the range of $[0.1, 1]$, and can be tuned to obtain a best fit.

\vspace{-5pt}
\section*{Acknowledgement}
\vspace{-5pt}
The authors acknowledge support from SRC 
(Grant SRC
2021-LM-3042) and NSF
(Grant ECCS 16-53241 CAR).

\vspace{-10pt}
\bibliographystyle{IEEEtran}
\bibliography{references}

\end{document}